# Trap with ultracold neutrons as a detector of dark matter particles with long-range forces

A.P. Serebrov[*], O.M. Zherebtsov
Petersburg Nuclear Physics Institute, Russian Academy of Sciences,
Gatchina, Leningrad District, 188300, Russia

Abstract

The possibility of using a trap with ultracold neutrons as a detector of dark matter particles with long-range forces is considered. The basic advantage of the proposed method lies in possibility of detecting the recoil energy $\sim 10^{-7}$ eV. The restrictions on parameters of type $\varphi(r) = a \dfrac{e^{-r/b}}{r}$ interaction potential between dark matter particles and a neutron are presented for different dark matter densities on the Earth. The assumption concerned with long-range interaction of dark matter particles and ordinary matter leads to a substantial enhancement of cross section at low energy. Consequently, there arises a possibility of capture and accumulation of dark matter in a gravitational field of the Earth. Rough estimation of accumulation of low-energy dark matter on the Earth is discussed. The first experimental restrictions for existence of dark matter with long-range forces on the Earth are presented.

## Introduction

The problem of dark matter in the Universe is one of the most important and the most interesting problems in modern physics [1]. The nature of dark matter is unknown, though there are some estimations on its quantity in the Universe [2] and on its location in galaxies [2]. These estimations point out that the dark matter density approximately 5 times [2] exceeds that of the ordinary baryon matter. Dark matter is invisible in the observable spectral range, but manifests itself in gravitational influence on movement of stars in galaxies, on movement of galaxies and hot gas clouds in congestions of galaxies which may be regarded as evidence for its existence [3,4]. For us it is significant two statements that can be ascertained reliably: 1) Interaction of dark matter with ordinary matter is very weak. For instance, when WIMP are considered as dark matter particles the cross section of interaction is less than $10^{-40}$ cm$^2$. This statement is based on the results of numerous experiments devoted to search for dark matter [2]. 2) Dark matter is localised in the so-called halo of galaxies and in congestions of galaxies. The latter is a widely known fact as the movement velocity of stars and of cold gas clouds in the Galaxy depending on the distance to the Galaxy centre [2] can not be interpreted without taking into account some additional invisible mass located in the Galaxy halo. From these studies it follows that the density of the galactic dark matter is $\rho_{\text{DM}}^{\text{gal}} \sim 0.3 \text{GeV/cm}^3 \approx 5 \cdot 10^{-25} \text{g/cm}^3$ [2].

Experiments on search for a dark matter are carried out in more than 20 laboratories of the world. Most of them are focused on direct registration of the dark matter particles with big energy or mass. Special approach is used in the DAMA/NaI, DAMA/LIBRA experiment where the time dependence on lower background part of the spectrum in the area of 2-6 keV is detected

---

[*] E-mail; serebrov@pnpi.spb.ru
Phone : +7 81371 46001
Fax : +7 81371 30072

[5]. The annual background variations are considered to be criterion of a dark matter signal. Such a signal is found in the experiment that has been carried out for 10 years. Variation of the background signal with an annual period and its maximum on the 2nd of June is observed to be reliable at ~8.9 standard deviations. The signal of such character is supposed to be due to the movement of the solar system through the halo of the Galaxy at the velocity of 230 km/s and the Earth movement round the Sun at the velocity of 30 km/s. It results in emerging variations of the dark matter flux particles passing through a detector located on the Earth. The exposition in the DAMA/NaI + DAMA/LIBRA experiment makes up $2.5 \cdot 10^5$ kg×day, exceeding that in other experiments by a few orders of magnitude. Another distinguished feature of the DAMA detector is a low threshold of signal registration equal to 2 keV. The energy spectrum of signals changing annually is presented in Fig. 1 from the work [5].

The registered energy is observed to be reducing as the detector count rate increases. The further course of the spectrum in the area below the threshold of 2 keV is obscure. One can assume the further growth of the count rate of events as energy decreases. Such an increase of the count rate is likely to be caused by interaction of dark matter particle with long-range forces. In this case the elastic cross section at small recoil energies grows in accordance with the well-known dependence: $\frac{d\sigma}{d\varepsilon} \sim \frac{1}{E_R^2 v_{DM.}^2}$, where $E_R$ - the recoil energy, $v_{DM.}$ - velocity of a dark matter particle.

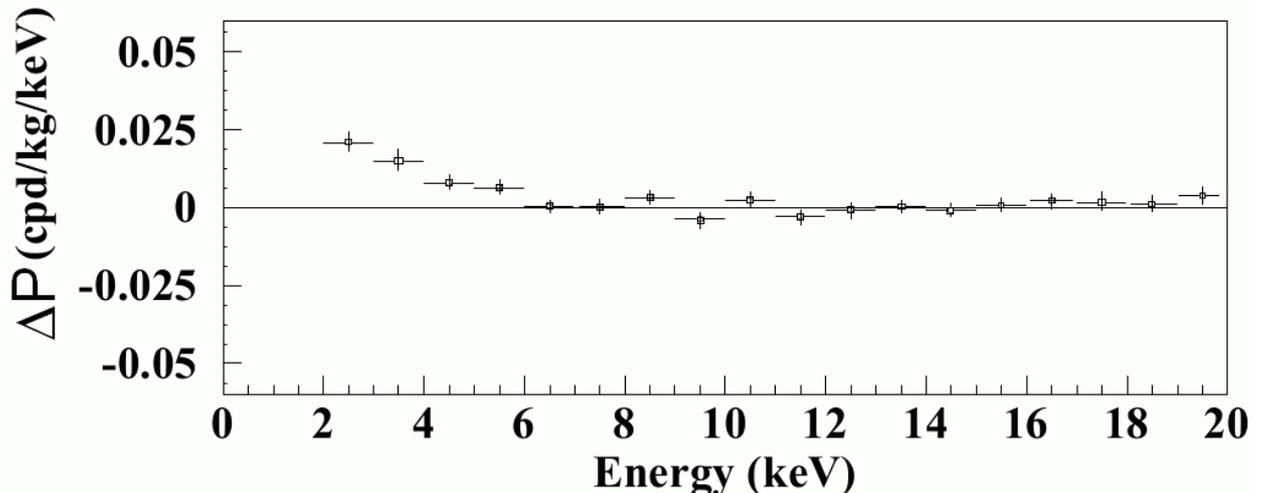

Fig.1 Variation part of energy spectrum of signals with the annual period in the experiment DAMA.

In agreement with the above made assumptions it is desirable to raise the question, formulated in the title of the following paragraph.
In further considerations we will regard the observed signal in the DAMA experiment to be concerned with a dark matter though such a statement is not yet proved.

## 1. Can a trap with ultracold neutrons be a detector of dark matter with long-range forces?

At first sight the answer to this question is negative since the density of ultracold neutrons in a trap is by 20 orders of magnitude less than that of atoms in substance. However, the registration threshold of dark matter by means of such a detector will be by 10 orders of magnitude lower, than that in the DAMA experiment. The point is that as soon as the ultracold neutron obtains the recoil energy of $10^{-7}$ eV it will immediately leave a trap, with this event being detected.

Let us make estimations of UCN elastic scattering cross section and the dark matter particles possessing long-range forces. The interaction potential is to be presented in the most general phenomenological form $\varphi(r) = a \dfrac{e^{-r/b}}{r}$.

However, one can somewhat specify the idea involved, keeping in mind that one deals with the modified gravitation at short distances [6] $\varphi(r) = -\dfrac{\alpha G_N m_n^2 M_{DM}}{r} e^{-r/\lambda}$, where $G_N$ is a gravitational constant, $\lambda$ is an effective radius of potential, $\alpha$ is a parameter of modified gravitation, $m_n$ - a nucleon mass, $M_{DM}$ - a dark matter particle mass presented in units of a neutron mass. It is worth considering the interaction by means of an axion-like particle exchange with a small mass of about $10^{-3}$-$10^{-5}$ eV providing the radius of forces corresponds to $10^{-3}$-1 cm : $\varphi(r) = -\dfrac{g_{aOM} g_{aDM} M_{DM}}{4\pi r} e^{-r/\lambda}$, where $g_{aOM}$ is a coupling constant of the spin independent interaction between an axion-like particle and a nucleon, and $g_{aDM}$ is a coupling constant of the spin independent interaction between an axion-like particle and a dark matter particle. The above stated potential is constructed in such a way, that at one of vertices an axion-like particle interacts with ordinary matter but at another one interacts with DM, so we are to substitute $g_{aOM} g_{aDM}$ instead of $g_{aOM}^2$. A relation of $\alpha$ and $g_{aOM} g_{aDM}$ can be written by analogy with known parity ($g_{aOM}^2 = \alpha 4\pi G_N m_n^2$), namely $g_{aOM} g_{aDM} = \alpha 4\pi G_N m_n^2$.

Finally, it is possible to treat the idea of kinematic mixing of a mirror photon and an ordinary one discussed in the work [7], concerned with mirror matter – an alternative for dark matter. Certainly, in our case the interaction of dark mirror matter with neutron is likely to occur due to the neutron magnetic moment, which will give rise to an additional small value.

However, one can write for discussion of experiment DAMA: $\varphi(r) = \dfrac{\varepsilon_{mirror} G_q zZ}{r} e^{-r/\lambda}$ , , where $\varepsilon_{mirror}$ -a mixing constant between a mirror photon and an ordinary one, $G_q$ - a constant of Coulomb interaction, $z$ -a charge of the target atoms, $Z$ - a charge of the dark matter particles. This idea has been used for the interpretation of the DAMA experiment results [7].

At last it should be mentioned about rather interesting review A.D. Dolgov (1999) article devoted to long-range forces in the Universe [8], where wide spectrum of ideas have been analysed.

New theoretical models of dark matter particles consider in detail in [9-11].

Let us make an estimation of UCN scattering cross-section with a dark matter particle possessing long-range interaction. To be specific, we take the modified gravitation (2). Generalisation on the interaction by axion-like particle exchange (3) can be made using a ratio of constants $\alpha$ and $g_{aOM} g_{aDM}$. In the Born approximation the scattering amplitude connected with a modified gravitational interaction is easily calculated by the formula from [12]: $f = -\dfrac{m}{2\pi\hbar^2} \int \varphi(r) e^{-i\vec{q}\vec{r}} dV$, where $q = |\vec{k}' - \vec{k}|$, $\vec{k}$ and $\vec{k}'-$ wave vectors of an originally based particle in the centre of mass system prior and post impact, $m$ -a reduced mass $m = \dfrac{m_1 m_2}{m_1 + m_2}$. Calculation $f$ in (5) yields $f = \dfrac{2m}{\hbar^2} a \dfrac{\lambda^2}{(\lambda q)^2 + 1}$, where coefficient $a$ may have three different value in dependece on considered case: $a = -\alpha G_N m_n^2 M_{DM}$, $a = -g_{aOM} g_{aDM} M_{DM}/4\pi$ or $a = \varepsilon_{mirror} G_q zZ$. In our calculations we take account of the following:

Firstly, the particle $m_1$ before colliding with the particle $m_2$ is at rest, thereby in the laboratory system of co-ordinates the particle $m_1$ after impact obtains a momentum $q$ which is related to the recoil energy - $\varepsilon$ by a simple equation $q = \frac{\sqrt{2\varepsilon m_1}}{\hbar}$. Secondly, an angular distribution in the centre of mass system is isotropic, i.e. all angles $\theta$ are equiprobable. Thus the following relation from [13] holds true for: $q = 2mv_{DM}\sin(\theta/2)$, where $v_{DM}$ - velocity of a incident particle at infinity. From (8) we have the expression for $\sin(\theta/2)$, taking into consideration that $d\Omega = 2\pi\sin\theta d\theta = 8\pi\sin(\theta/2)d\sin(\theta/2)$, we get: $d\Omega = 8\pi\frac{qdq}{(2mv_{DM})^2}$. then, taking into account the relationship between an impulse $q$ and the energy, one can derive: $d\Omega = \frac{\pi(m_1+m_2)^2}{m_1 m_2}\frac{d\varepsilon}{E_{DM}}$, where $E_{DM}$ - energy of a incident particle at infinity. Having squared $f$ and multiplying it by $d\Omega$ one obtains that differential cross section of particle of mass $m_2 = M_{DM}$ on a mass $m_1 = M_{OM}$ particle can be expressed in the following way:

$$d\sigma = \frac{4\pi\alpha^2 G_N^2 m_n^6}{\hbar^4}\frac{\lambda^4(M_{OM}M_{DM})^3}{(2\lambda^2 m_n M_{OM}\varepsilon/\hbar^2 + 1)^2}\frac{d\varepsilon}{E_{DM}} = 1.26\cdot 10^{-48}\cdot\alpha^2\frac{\lambda^4(M_{OM}M_{DM})^3}{(5\cdot 10^{18}\lambda^2 M_{OM}\varepsilon + 1)^2}\frac{d\varepsilon}{E_{DM}}, \quad (1)$$

where cross section expressed in cm$^2$, $M_{OM}$ - mass of a scattered particle of ordinary matter, expressed in units of nucleon mass $m_n$, $M_{DM}$ - mass of an incident dark matter particle, expressed in units of nucleon mass $m_n$, $\varepsilon$ - the recoil energy expressed in electron-volts, a force action radius - $\lambda$ is expressed in centimetres. $E_{DM}$ - energy of an incident dark matter particle, $E_{DM}[eV] = M_{DM}\cdot 5.2\cdot 10^{-9}v_{DM}^2[m/s]$. Provided that $\lambda^2 m_1\varepsilon/\hbar^2 \gg 1$ expression (1) is simplified, it takes the following form: $d\sigma = \frac{\pi\alpha^2 G_N^2 m_n^4 M_{OM} M_{DM}^3}{E_{DM}}\frac{d\varepsilon}{\varepsilon^2}$. This formula is similar to that for the Rutherford scattering case.

For integrated cross section using formula (1) we have calculated:

$$\sigma_{DM-OM}(\varepsilon_R^{max}, \varepsilon_R^{min}) = \frac{4\pi G_N^2 m_n^6}{\hbar^4}\frac{(\varepsilon_R^{max} - \varepsilon_R^{min})}{E_{DM}}\frac{\lambda^4\alpha^2(M_{OM}M_{DM})^3}{(5\cdot 10^{18}\varepsilon_R^{max} M_{OM}\lambda^2 + 1)(5\cdot 10^{18}\varepsilon_R^{min} M_{OM}\lambda^2 + 1)}, \quad (2)$$

where $\varepsilon_R^{min}$ is minimum recoil energy, $\varepsilon_R^{max}$ is maximum recoil energy equal to $\varepsilon_R^{max} = \frac{4M_{OM}M_{DM}}{(M_{OM}+M_{DM})^2}E_{DM}$.

In our case of scattering a dark matter particle on a neutron it is reasonable to use the velocity of 8 km/s (the 1$^{st}$ - orbital velocity) since we assume dark matter to be localised in circumterrestrial orbits near to the Earth surface.

Fig.2 a) illustrates the integral cross section of UCN as a function of the registration threshold for different $\lambda$ values. On the axis of ordinates the cross section values divided by $\alpha^2$ are presented. Fig.2b) shows a graph taking into account experimental conditions of the DAMA experiment, the only difference being that the recoil nucleus mass is equal to 23 nucleon mass (NA), the dark matter particle velocity is 230 km/s. On the left axis of ordinates the cross section values divided by $\alpha^2$ are presented, while on the right axis the cross section values divided by $\varepsilon_{mirror}^2$ are given. The curve shows the cross section scale according to the choice of the interaction force. It is quite obvious that gravitational interaction at short distances must be considerably intensified to make experimental observations by means of UCN ($\alpha \gg 1$). On the other hand, the interaction is supposed to be much less than the Coulomb one ($\varepsilon_{mirror} \ll 1$) since magnitudes on the right axis for $\varepsilon_R^{min} \approx 10^{-7}$ eV and $\lambda \approx 10^{-4}$ cm are abnormally big ($10^{20}$ barn). Making comparison of cross section for UCN at threshold of $10^{-7}$eV with that for the recoil nucleus in the DAMA experiment at threshold of 2 keV one can conclude that the cross section increase factor equals 10 orders of magnitude. Here we take both parametre $\alpha$ and the dark matter particle mass (20 nucleon units) to be the same as those in the DAMA experiment.

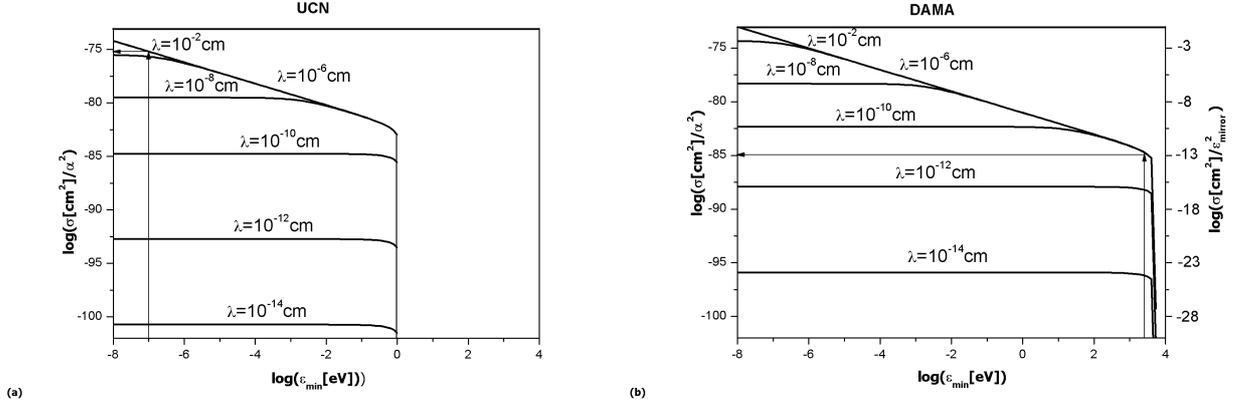

Fig.2 a) Total cross section of the elastic scattering of UCN depending on the minimum recoil energy for the different $\lambda$ values. Mass of a dark matter particle is 20 nucleon mass, velocity is 8 km/s, at the energy threshold being equal to $10^{-7}$eV $\sigma = \alpha^2 10^{-75}[\text{cm}^2]$. b) cross section of the elastic scattering of Na depending on the minimum recoil energy for different $\lambda$ values. Mass of a dark matter particle is 20 nucleon mass, velocity is 230 km/s, at the energy of threshold being equal to 2 keV $\sigma = \alpha^2 10^{-85}[\text{cm}^2]$.

## 2. The analysis of DAMA experiment

The analysis of the DAMA experiment has been made on density of galactic dark matter equal to 0.3 GeV/cm$^3$, velocity of dark matter particles equal to 230 km/s. The maximum recoil energy in this experiment is approximately 6 keV (Fig. 1). Atoms Na (A=23) and I (A=127) are taken as a recoil nucleus. Using formula for the maximal recoil energy and from the listed data one receives mass of dark matter particle. Recoil energy of 6 keV for sodium atom (A=23) is obtained at mass of a scattering particle of 20 GeV, while for iodine atom (A=127) it is obtained at mass of a scattering particle of 32 GeV. Further for simplicity we will be using mass of 20 GeV.

The probability of registration of a dark matter particle by a nucleus of Na is the following:

$$P(t) = n_{DM} v(t) \sigma_{DM} = n_{DM} \sigma_{DM} \left( v_{Sun} + v_{Earth} \cos 60^0 \cos(2\pi(t-t_0)/T) \right), \quad (3)$$

where density of dark matter - $n_{DM} = \dfrac{0.3}{20} = 1.5 \cdot 10^{-2}$ cm$^{-3}$, $v(t)$ - detector velocity, $v_{Sun}$ -the Sun velocity 230 km/s, $v_{Earth}$ - velocity of the Earth around the Sun 30 km/s, the slope angle of terrestrial orbit plane towards the direction of the Sun movement equals $60^0$, $\sigma_{DM}$ - cross section for scattering dark matter particle with nucleus of the detector. Thus, variation of registration probability is:

$$\Delta P(t) = n \, \sigma_{DM} (v_{Sun}) v_{Sun} 0.07 \cos(2\pi(t-t_0)/T). \quad (4)$$

Fig. 1 shows the variation amplitude of the detector count rate to be 0.02 events $\times$days$^{-1} \times$ kg$^{-1}$ $\times$keV$^{-1}$ at 2 keV, the integrated value on a spectrum from 2 keV to 6 keV equal to $4 \cdot 10^{-2}$ days$^{-1}$ $\times$kg$^{-1}$. Using the following values from formula (4): $n_{DM} = 1.5 \cdot 10^{-2}$ cm$^{-3}$; days $= 8.6 \cdot 10^4$ s; 1kg $\sim 4 \cdot 10^{24}$ atoms Na, recoil cross section of Na is obtained: $\sigma_{DM} = 4.75 \cdot 10^{-36}$ cm$^2$. Then, from formula (2) we have the ratio:

$$4.75 \cdot 10^{-36} = 1.26 \cdot 10^{-48} \frac{\alpha^2 \lambda^4 (M_{OM} M_{DM})^3}{(5 \cdot 10^{18} \varepsilon_R^{max} M_{OM} \lambda^2 + 1)(5 \cdot 10^{18} \varepsilon_R^{min} M_{OM} \lambda^2 + 1) E_{DM}}, \quad (5)$$

where $\varepsilon_R^{max} \approx 6$ keV, as one can see from Fig.1 $\varepsilon_R^{min} = 2$keV is the registration threshold.

Relationship between values $\alpha$ and $\lambda$ derived by (5) is presented in Fig.3. From Fig.3 one can conclude, that the area of values $\alpha$ is defined within $10^{26}$-$10^{28}$. Because of registration of events in the DAMA experiment below a threshold 2keV is impossible, the parameter $\lambda$ is not defined by anything. Consequently, the assumption about possibility of long range type interaction of dark matter particles with ordinary matter is not closed. But if it exists, then $\alpha \approx 10^{26}$. However, the area of $\lambda$ values practically is not defined due to rather high experimental threshold 2 keV. Value of $\lambda$ could be defined if present hypothetical possibility of decreasing of experimental threshold. In this case the differential cross section (counting rate of events in Fig.1) be to increase until achieve the value defined by a parity $5 \cdot 10^{18} \varepsilon_R^{min} M_{OM} \lambda^2 \approx 1$ (see the formulae (1) and (2)). Then $\lambda$ is defined from the following expression $\lambda = (5 \cdot 10^{18} \varepsilon_R^{min} M_{OM})^{-1/2}$, where $\varepsilon_R^{min}$ - value of a detector threshold at which growth of the counting rate stops.

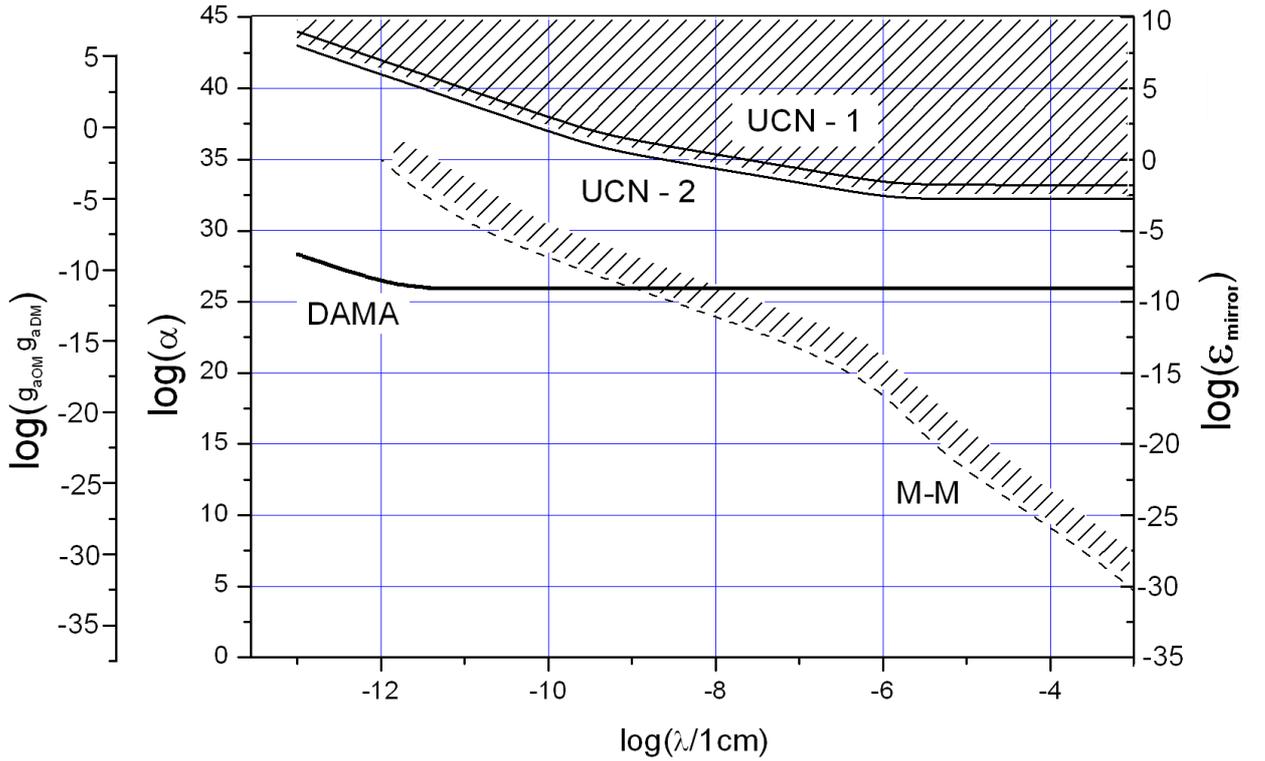

Fig. 3 Area of the parameters $\alpha$ and $\lambda$ (or $g_{aOM} g_{aDM}$ and $\lambda$) from DAMA experiment. The area of constraints UCN-1 is obtained from the condition $(\tau_n^{-1})_{UCN} - (\tau_n^{-1})_{beam} \leq 2 \cdot 10^{-5} s^{-1}$, product $k_{DM}^{S-E} \cdot k_{DM}^{E-M} \cdot k_{DM}^{E-surface}$ is taken equal to 1 The area of constraints UCN-2 is obtained from the limit on probability of low-upscattering of UCN $(\tau_{up}^{-1})_{DM} \leq 3 \cdot 10^{-7} s^{-1}$ [14], product $k_{DM}^{S-E} \cdot k_{DM}^{E-M} \cdot k_{DM}^{E-surface}$ is taken equal to 1. Area M-M corresponds to constraints on interactions of an ordinary matter with an ordinary matter [8].

# 3. Limits on parametres $\alpha$ and $\lambda$ from the experiment with ultracold neutrons

Let us estimate the limits on observability of the parameters $\alpha$ and $\lambda$ in the experiment with a UCN trap as a detector of dark matter. Consideration can be started with a rough estimate, using the following fact. Measurement of neutron lifetime by means of UCN might contain a systematic error because of interaction of UCN with dark matter.

The probability of UCN being stored in the trap involves the following factors:

$$\tau_{st}^{-1} = \tau_n^{-1} + \tau_{loss}^{-1} + \tau_{vac}^{-1} + \tau_{DM}^{-1}, \quad (6)$$

where $\tau_n^{-1}$ - probability of decay, $\tau_{loss}^{-1}$ - probability of losses because of collisions with trap walls, $\tau_{vac}^{-1}$ - probability of losses because of residual vacuum, $\tau_{DM}^{-1}$ - probability of losses because of scattering by dark matter particles.

Measurement of neutron lifetime on a neutron beam by registering decay products will not contain such a systematic error. Therefore we are able to make an estimation on the upper limit of the probability of UCN and dark matter interaction.

$$\left(\tau_n^{-1}\right)_{UCN} - \left(\tau_n^{-1}\right)_{beam} = \tau_{DM}^{-1} = (1.10 \pm 0.45) \cdot 10^{-5} s^{-1} < 2 \cdot 10^{-5} s^{-1}, \quad (7)$$

where $\left(\tau_n\right)_{UCN}$ is neutron lifetime measured by the UCN method (the most accurate result of measurements: $878.5 \pm 0.8$ s [15]), $\left(\tau_n\right)_{beam}$ is neutron lifetime measured on a neutron beam by registering decay products of neutron (the most accurate result of measurements: $886.8 \pm 3.4$ s [16]). Comparing the measured results we conclude that the difference does not exceed 2% with the reliability level of 95%.

Now we can write the following equation: $n_{DM} \sigma_{DM\text{-}UCN} v_{DM} \leq 2 \cdot 10^{-5} s^{-1}$, where $n_{DM}$ is the density of dark matter particles, $v_{DM}$ is the velocity of dark matter particles, $\sigma_{UCN\text{-}DM}$ is the elastic cross section of UCN and a dark matter particle. Formula (2) will be used for the interaction cross section. We will assume dark matter particles to be on the circumterraneous orbits and to have the velocity of 8 km/s. The most complicated question is concerned with the density of dark matter particles being on circumterraneous orbits. In accordance with the works [17,18] dark matter density on the orbit of the Earth around the Sun ($\rho_{DM}^{S\text{-}E}$) $10^3$-$10^4$ times exceeds the galactic density - $\rho_{DM}^{gal}$. However, in a number of works [19-21] is shown that density of the captured dark matter of the same order, as galactic density of a dark matter. In above mentioned works long range interaction was not considered. Below we will discuss that owing to growth of cross section at low energies account of collision processes in substance of an gravitational object would lead to losse of energy by a dark matter particle and to considerable factors of dark matter capture in a gravitational hole of the object. While such calculations are not executed. Nevertheless we can introduce corresponding factors of capture with aim to get restrictions on their value from experiments with UCN. We will notice that we are going to consider factors of low-energy dark matter capture.

Let us determine the factors of low energy dark matter capture. The factor of dark matter capture on the orbit of Earth is equal to $k_{capt}^{S\text{-}E} = \dfrac{\rho_{DM}^{S\text{-}E}}{\rho_{DM}^{gal}}$. The factor of dark matter capture in the system Earth-Moon is $k_{capt}^{E\text{-}M} = \dfrac{\rho_{DM}^{E\text{-}M}}{\rho_{DM}^{S\text{-}E}}$. At last, it is necessary to introduce the factor of the dark matter density increasing on the surface of the Earth in relation to the average value of density in the system Earth-Moon - $k_{capt}^{E\text{-}surface} = \dfrac{\rho_{DM}^{E\text{-}surface}}{\rho_{DM}^{E\text{-}M}}$. We have to consider as a necessary condition that in case of increasing mass of dark matter particles, the quantity of particles of dark matter in cm$^3$ - $n_{DM}$ decreases, keeping average value of Galactic density

$\rho_{DM}^{gal} = 0.3$ GeV/cm$^3$, $n_{DM} = \frac{\rho_{DM}^{gal}}{M_{DM}}$. However, the factors $k_{capt}^{S-E}$, $k_{capt}^{E-M}$ and $k_{capt}^{E-surface}$ are independent of the mass of dark matter particles. Thus, $n_{DM}$ equals to: $n_{DM} = \frac{\rho_{DM}^{gal}}{M_{DM}} k_{DM}^{S-E} k_{DM}^{E-M} k_{DM}^{E-surface}$.

Let's notice that capture factors are connected with energy of dark matter particles which can be kept by a gravitational field of corresponding objects. For the analysis of DAMA experiment speed of dark matter particles 230 km/s has been used. It means that in calculations the galactic density without additional capture factors is used. Capture factors discussed above concern to dark matter particles energies being considerably below a threshold 2 keV as in the DAMA experiment. But these factors of capture have the complete relation to UCN experiment with, where a registration threshold is very low. Let's consider what restrictions can be derived from the equation (7).

The restriction $\tau_{DM}^{-1} < 2 \cdot 10^{-5}$ c$^{-1}$ yields the following equation:

$$7.27 \cdot 10^{-39} k_{DM}^{S-E} \cdot k_{DM}^{E-M} \cdot k_{DM}^{E-surface} \cdot \frac{\left(\varepsilon_R^{max} - \varepsilon_R^{min}\right)}{v_{DM}} \frac{M_{DM} \lambda^4 \alpha^2}{\left(5 \cdot 10^{18} \varepsilon_R^{max} \lambda^2 + 1\right)\left(5 \cdot 10^{18} \varepsilon_R^{min} \lambda^2 + 1\right)} \leq 2 \cdot 10^{-5} s^{-1}, (8)$$

where $\varepsilon_R^{max}, \varepsilon_R^{min}$ are expressed in eV, $\lambda$ - in cm, $M_{DM}$ - in nucleon mass, velocity $v_{DM}$ - in m/s. From the condition $5 \cdot 10^{18} \varepsilon_R^{min} \lambda^2 \gg 1$, we derive a limit on $\alpha$ at $\lambda$ of about 10$^{-5}$ cm:

$$\alpha \leq 5.8 \cdot 10^{31} \cdot \sqrt{\frac{v_{DM}}{k_{DM}^{S-E} \cdot k_{DM}^{E-M} \cdot k_{DM}^{E-surface} M_{DM}}}. \quad (9)$$

Fig. 3 shows the area of restrictions of the parametres $\alpha$ and $\lambda$ (or $g_{aOM} g_{aDM}$ and $\lambda$) from the condition $\left(\tau_n^{-1}\right)_{UCN} - \left(\tau_n^{-1}\right)_{beam} \leq 2 \cdot 10^{-5} s^{-1}$. Here is also presented the range of parameter values $\alpha$ and $\lambda$ (or $g_{aOM} g_{aDM}$ and $\lambda$) from the DAMA experiment. The data analysis of the DAMA experiment is made within the frame of the same phenomenological model. As the registration threshold in the DAMA experiment is limited to 2 keV, nothing can be said about $\alpha$-value at $\lambda > 10^{-12}$ cm. Thus, for large $\lambda$ in the DAMA experiment $\alpha$ value is determined at the level of 10$^{26}$ or $g_{aOM} g_{aDM}$ is determined at the level of 10$^{-11}$. In Fig. 3 the restriction curve UCN-1 has been calculated with the same mass of dark matter particles as in the DAMA experiment, i.e. equal to 20 nucleon mass. One can see that at $\lambda \sim 10^{-4}$ cm the restrictions on the parametres $\alpha$ or $g_{aOM} g_{aDM}$ are by 6 order worse, than values $\alpha$ or $g_{aOM} g_{aDM}$ from the DAMA experiment. However, these restrictions do not take into account the factors of capture $k_{DM}^{S-E}, k_{DM}^{E-M}$ and $k_{DM}^{E-surface}$. We can make the maximum estimation on the product of all three factors $k_{DM}^{S-E} \cdot k_{DM}^{E-M} \cdot k_{DM}^{E-surface}$, assuming that the dark matter signal in the DAMA experiment is really registered, dark matter having a long-range interaction $\lambda \approx 10^{-5} - 10^{-3}$ cm. Then equating $\alpha$ in formula (9) to the estimation $\alpha = 10^{26}$ from the DAMA experiment, we have the upper limit on the product of factors: $k_{DM}^{S-E} \cdot k_{DM}^{E-M} \cdot k_{DM}^{E-surface} \leq 1.3 \cdot 10^{14}$. Accordingly, limits on density of long range dark matter on the Earth are the following: $\rho_{DM}^{E-surface} \leq 6.6 \cdot 10^{-11}$ g/cm$^3$. The value $k_{DM}^{S-E} \cdot k_{DM}^{E-M} \cdot k_{DM}^{E-surface} > 1.3 \cdot 10^{14}$ does not meet the condition $\left(\tau_n^{-1}\right)_{UCN} - \left(\tau_n^{-1}\right)_{beam} = \tau_{DM}^{-1} \leq 2 \cdot 10^{-5} s^{-1}$, i.e. neutron lifetime in the experiment with UCN would be considerably suppressed.

In Fig. 3 area M-M corresponds to restrictions on deviations from gravitational interaction at short distances for interactions of ordinary matter with ordinary matter [6]. In area $\lambda = 10^{-12}$ cm value $\alpha = 10^{26}$ from DAMA experiment does not contradict M-M restrictions. If one transfers this value $\alpha$ for area $\lambda \simeq 10^{-5}$ cm, one turns out to be in the area excluded for interaction of ordinary matter with ordinary matter. However, we take interest in the interaction of dark matter with an ordinary one. Therefore, there remains possibility to discuss the theory of exhange of an axion-like particle having different coupling constants of this particle with ordinary and dark matter. From a relation $g_{aOM}^2 = 10^{-37} \times \alpha$ and experimental restriction $\alpha \leq 10^{15}$ for $\lambda \approx 10^{-6} - 10^{-5}$ cm it follows that $g_{aOM} < 10^{-11}$. Then, $g_{aDM} \approx 1$. For monopole-dipole interaction the experimental limits are much weaker $g_s g_p < 3 \cdot 10^{-12}$ for $\lambda \approx 10^{-5}$ cm [22,23]. Summarizing, one should say that since the models of interaction of dark and ordinary matter are practically absent, the modelless experimental limits are of their own value.

The first impression about the possibility of using a UCN trap for detecting dark matter possessing long range radius of force on the Earth is not optimistic. However, in the case of long range interaction factors of low energy dark matter capture by the Earth-Moon system have not been determined in any way for the time being. Moving along the orbit in the solar system where the density of dark matter can be increased approximately by four orders of magnitude in relation to the galactic density, the system Earth-Moon could have saved up a significant amount of dark matter during its existence. Besides, the density of dark matter near the Earth surface is likely to be higher than an average density in the system Earth-Moon.

The maximum estimation of the dark matter mass on the Earth has been made in the work [24] in terms of possible influence of dark matter on geophysical characteristics of the Earth. The maximum quantity of dark matter on the Earth makes up not more than $3.8 \cdot 10^{-3}$ of the Earth mass, i.e. the upper limit of dark matter on the Earth is $10^{-5} - 10^{-3}$ g/cm$^3$. Undoubtedly, this value is extremely high and it would be not reasonable to use it in the calculations, since according to our estimations the maximum density is $\rho_{DM}^{E\text{-surface}} \leq 6.6 \cdot 10^{-13}$ g/cm$^3$.

Now it should be noticed that sensitivity estimation of the experiment with a UCN trap aimed at searching for dark matter is rather rough. The matter is that the estimation is made comparing neutron lifetime measured by different methods and makes 1%. Here the absolute measurements of neutron lifetime have been used. If variations of UCN storage time are investigated, the measurements can be made within accuracy of 0.01 % as these are relative measurements. In this case the sensitivity of the experiment for determination of $\alpha$ will be:
$\alpha_{\text{sensitivity}} \approx 6 \cdot 10^{30} \cdot \sqrt{\dfrac{v_{DM}}{k_{DM}^{S\text{-}E} \cdot k_{DM}^{E\text{-}M} \cdot k_{DM}^{E\text{-surface}} M_{DM}}}$. However, one should keep in mind that at small depth of the variation the sensitivity of the UCN method falls.

From this formula it follows that the experiment is rather sensitive to detecting big masses with a long-range interaction. For example, for the mass of dark matter particles $M_{DM} = 10^{10}$ nucleon mass : $\alpha_{\text{sensitivity}} \approx 6 \cdot 10^{25} \cdot \sqrt{\dfrac{v_{DM}}{k_{DM}^{S\text{-}E} \cdot k_{DM}^{E\text{-}M} \cdot k_{DM}^{E\text{-surface}}}}$. Probably, the UCN method will allow super heavy particles of dark matter to be detected. However, this conclusion is valid only if the interaction constant is proportional to mass.

To conclude, it should be mentioned that as to the experimental problem of detecting the recoil energy of dark matter particles possessing the kinetic energy of 5-10 eV, it is rather difficult to propose the method for detecting energies lower than 1 eV. In this regard the UCN method is a unique one.

## 4. An experimental estimation of a limit on effect of heating UCN by scattering dark matter particles

The purpose of the experiment [14] with UCN was to study carefully the effect of low energy heating (increasing of kinetic energy) UCN at reflection from trap walls. The effect of heating has been found to be $(2.2 \pm 0.2)10^{-8}$ per one collision with a trap wall. It is worth using this result for the estimation of the upper limit on heating UCN by interaction with dark matter particles. In this experiment the frequency of UCN collisions with a trap was 15 Hz. Thus heating probability is less than $3 \cdot 10^{-7} s^{-1}$. For estimation of heating probability UCN by dark matter particles the following equation can be used: $P_{up.DM} = n_{DM} \cdot \sigma_{DM}^{up} \cdot v_{DM} < 3 \cdot 10^{-7} s^{-1}$. In this formula one should use cross section of UCN upscattering with recoil energy transition from $0.5 \cdot 10^{-7}$ eV to $10^{-7}$ eV, as it was made in experiment [14]. The estimation of this cross section and the differential cross section dependent on transferred energy can be as follows:

$$\frac{d\sigma}{d\varepsilon_R} = \frac{4\pi G_N^2 m_n^6}{\hbar^4} \frac{\alpha^2 M_{DM}^3}{E_{DM}(5 \cdot 10^{18})^2 \varepsilon_R^2} = 5.04 \cdot 10^{-86} \frac{\alpha^2}{E_{DM}[\text{eV}]} \frac{M_{DM}^3}{\varepsilon_R^2[\text{eV}]}. \quad (10)$$

Then, scattering cross section with energy transition from $0.5 \cdot 10^{-7}$ eV to $10^{-7}$ eV is equal to: $\sigma(10^{-7}, 5 \cdot 10^{-8}) = 6 \cdot 10^{-76} \alpha^2$ cm$^2$. In this estimation we used as earlier $M_{DM} = 20$ [GeV], $m_n$ - neutron mass, energies $\varepsilon_R$ and $E_{DM}$ are expressed in eV.

It is possible to make restrictions for $\alpha$ at $\lambda \approx 10^{-5}$ cm from formula (9): $\alpha < 2 \cdot 10^{32}$. The corresponding area of limits on parametres $\alpha$ and $\lambda$ is shown in Fig. 3. It is approximately by one order better than the previous one from the condition $(\tau_n^{-1})_{UCN} - (\tau_n^{-1})_{beam} \le 2 \cdot 10^{-5} s^{-1}$. These limits have been made without taking account of possible density increase of dark matter on the Earth surface. Accordingly, the estimation on the maximum density of dark matter on the Earth surface is less by two orders of magnitude. $\rho_{DM}^{E\text{-surface}} \le 6.6 \cdot 10^{-13}$ g/cm$^3$.

## 5. Interaction of particles of a dark matter with atmosphere and the Earth substance

The assumption on the long range forces acting at distance $\sim 10^{-5}$ cm can lead to a substantial enhancement of cross section and decrease of penetration depth of dark matter particles in substance. As a result, observations in the underground experiments including the DAMA experiment will become impossible. However, there is nothing contradictictory about it. Let us make the required estimations. From the formula for differential cross section (11) it is possible to obtain a differential equation for energy losses by a dark matter particle in the atmosphere and the Earth substance:

$$\frac{dE}{dl} = n_{OM} \int_{E_{th}}^{E_{max}} \frac{d\sigma}{d\varepsilon_R} \varepsilon_R d\varepsilon_R = \frac{5.04 \cdot 10^{-86}}{E_{DM}} n_{OM} \alpha^2 M_{DM}^3 M_{OM} \ln\frac{E_{max}}{E_{th}}, \quad (11)$$

where $E_{max}$ - maximum recoil energy $E_{th}$ - thermal energy of substance (in (11) energy is expressed in eV, path length $l$ is expressed in cm, density of ordinary matter $n_{OM}$ is expressed in cm$^{-3}$). It is reasonable to stop integration at thermal energy of substance - $E_{th}$. From here one can derive the path length of a dark matter particle when its energy decreases $e$ times.

$$L_e[\text{cm}] = 8.6 \cdot 10^{84} E_{DM}^2 [\text{eV}] / n_{OM} \alpha^2 M_{DM}^3 M_{OM} \ln(E_{DM}/E_{th}). \quad (12)$$

Thus the following estimations for different substances and velocities can be made at $\alpha = 10^{26}$ and $\lambda = 10^{-5}$ cm. (The particle density in the atmosphere is equal to $n_{atmosp} = 2.5 \cdot 10^{19}$ cm$^{-3}$. The particle density in the Earth substance equals $n_{Earth} = 7.5 \cdot 10^{22}$ cm$^{-3}$).

$L_{Earth}(v_{DM} = 230 \text{km/s}) = 1.23 \cdot 10^{6} \text{km}$

$L_{atmosp}(v_{DM} = 8 \text{km/s}) = 1.2 \cdot 10^{5} \text{km}$

$L_{Earth}(v_{DM} = 8 \text{km/s}) = 0.4 \cdot 10^{2} \text{km}$

The atomic mass of ordinary substance in these estimations is taken to be equal to 30 nucleon mass. Using the estimations performed above for particle path length the conclusion is made that dark matter particles pass through the Earth at velocity of 230 km/s almost free of loss. The relaxation time for a particle movement at velocity of 8 km/s along the orbit over the Earth surface is approximately 5 hours, however at the height of about 50 km the relaxation time equals approximately 0.5 year.

Observation over Earth rotation speed around own axis showed that its relative decreasing is order by $2 \cdot 10^{-10}$ per year [25]. Assume conservation of momentum quantity and use the restriction on decreasing of speed of the Earth rotation one concludes that relative increasing of Earth mass to be less than by $2 \cdot 10^{-10}$ per year. From the restriction on time of dark matter accretion, we obtain the upper limit on density of dark matter in the Earth atmosphere: $\rho \leq 8 \cdot 10^{-7}$ g/cm$^3$ $= 1.6 \cdot 10^{18} \rho_{gal}$.

This estimation does not exlude the earlier made estimation on the maximum density of a dark matter (32) ($\rho \leq 6.6 \cdot 10^{-13}$ g/cm$^3$), which has been obtained from the assumption of long-range forces $\alpha \approx 10^{26}$ from the DAMA experiment as well as from the following condition $\left(\tau_{up}^{-1}\right)_{DM} \leq 3 \cdot 10^{-7} s^{-1}$ in the experiment on lower energy heating of UCN.

Another estimation can be made in terms of our estimation made above of the maximum density and the rate of dark matter accretion by the Earth. The rate of the Earth mass increase per year is: $\frac{\Delta M}{M} < 1 \cdot 10^{-13}$ year$^{-1}$, i.e. the Earth might have increased its mass by 0.1 % for 5 billion years.

On the whole, the estimations made lead to the conclusion, which doesn't seem to be contradictory, concerning the possibility of dark matter thermalization in the internal structure of the Earth and its accumulation there.

Finally, it is to be mentioned that if the temperature in the internal structure of the Earth rises up to $8 \cdot 10^{3}$ K, the dark matter particles with mass 1 GeV are expected to reach the velocity of
11.5 km/s at thermalization and are likely to escape to the solar system. Accordingly, this process is similar to that for the Sun, however, being more intensive.

It is of interest to analyse how the cross section increase of interaction with the dark matter discussed above can affect the ineterpretation of movement anomaly of space vehicles [26].

Now for the sake of comparison it is possible to consider energy losses by a dark matter particle if interaction is of short-range, for example, at $\lambda = 10^{-12}$ cm. Formula (1) yields:

$$\frac{d\sigma}{d\varepsilon_R} = 1.26 \cdot 10^{-48} \frac{\alpha^2 \lambda^4 (M_{OM} M_{DM})^3}{E_{DM}}. \quad (13)$$

From Fig. 3 and calculation of the DAMA experiment $\alpha = 10^{26}$ at $\lambda = 10^{-12}$ cm. Then, $\frac{d\sigma}{d\varepsilon_R} = \frac{2.72 \cdot 10^{-36}}{E_{DM}}$. For energy losses on length unit the following expression is used:

$$\frac{dE}{dl} = n_{OM} \cdot \frac{2.72 \cdot 10^{-36}}{E_{DM}} \int_0^{E_R^{max}} \varepsilon_R d\varepsilon_R = n_{OM} 1.36 \cdot 10^{-36} \left[ \frac{4 M_{OM} M_{DM}}{(M_{OM} + M_{DM})^2} \right]^2 E_{DM}, \quad (14)$$

for $M_{DM} = 20$ [GeV], $M_{OM} = 30$ [GeV] and $n_{OM} = 8 \cdot 10^{22}$ cm$^{-3}$

$$\frac{dE}{dl}[\text{eV/cm}] = 1.0 \cdot 10^{-13} E_{DM} [\text{eV}].$$

The path length of a dark matter particle before it loses its energy $e$ times is not dependant on the initial energy and is approximately equal to: $L_e = 1.0 \cdot 10^8$ km.

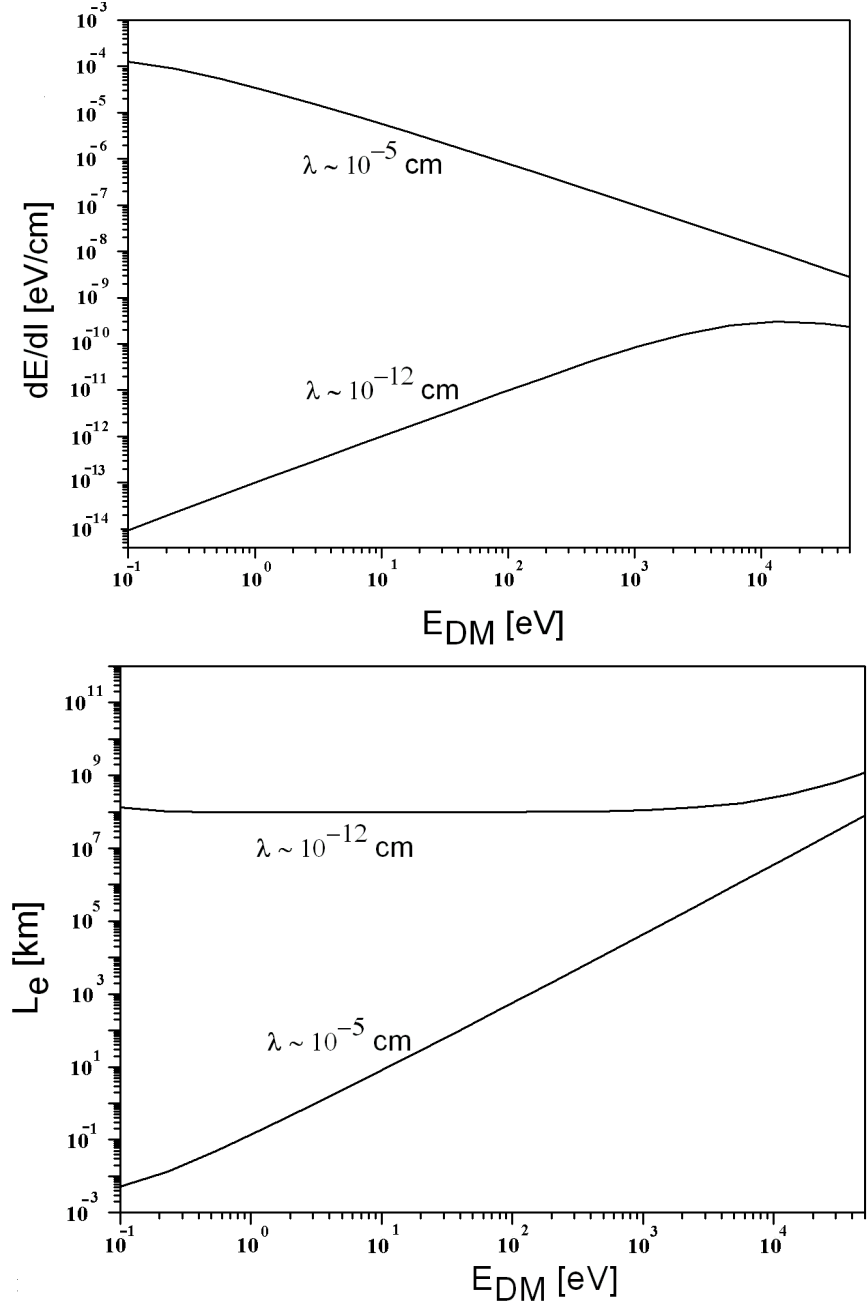

Fig. 4 Specific energy losses per unit of path and energy attenuation length as a function of dark matter particle energy for two cases $\lambda \approx 10^{-5}$ cm and $\lambda \approx 10^{-12}$ cm. ($\alpha = 10^{26}$, $n_{OM} = 8 \cdot 10^{22}$ cm$^{-1}$, $M_{DM} = 20$ [GeV], $M_{OM} = 30$ [GeV])

One can see that these two cases ($\lambda = 10^{-12}$ cm and $\lambda = 10^{-5}$ cm) differ by approximately 6 orders of magnitude for dark matter particle velocity equal to 10-11 km/s. Fig. 4 illustrates distinction between these cases within a wide range of energies. Such a difference results in important consequences. In case of short-range forces the dark matter capture in a gravitational hole of planets and the Sun is practically impossible [20,27]. Estimation of capture probability in a gravitational field of the Earth in the case of long range forces ($\lambda \approx 10^{-5}$ cm) is presented in the following paragraph.

It should be mentioned that at the low-energy effect of atom link in substance has to take place and effect of scattering by groupe of atoms due to a large value of the force action radius. These effects are not taken into account in given consideration.

## 6. An estimation of capture probability in a gravitational field of the Earth due to the interaction with the Earth substance

From the Fig. 5 it is clear that capture in a gravitational hole of the Earth will occur, if the energy $E_{DM\infty}$ (equal to kinetic energy of a particle at infinity) is dissipated due to the interaction with the Earth substance.

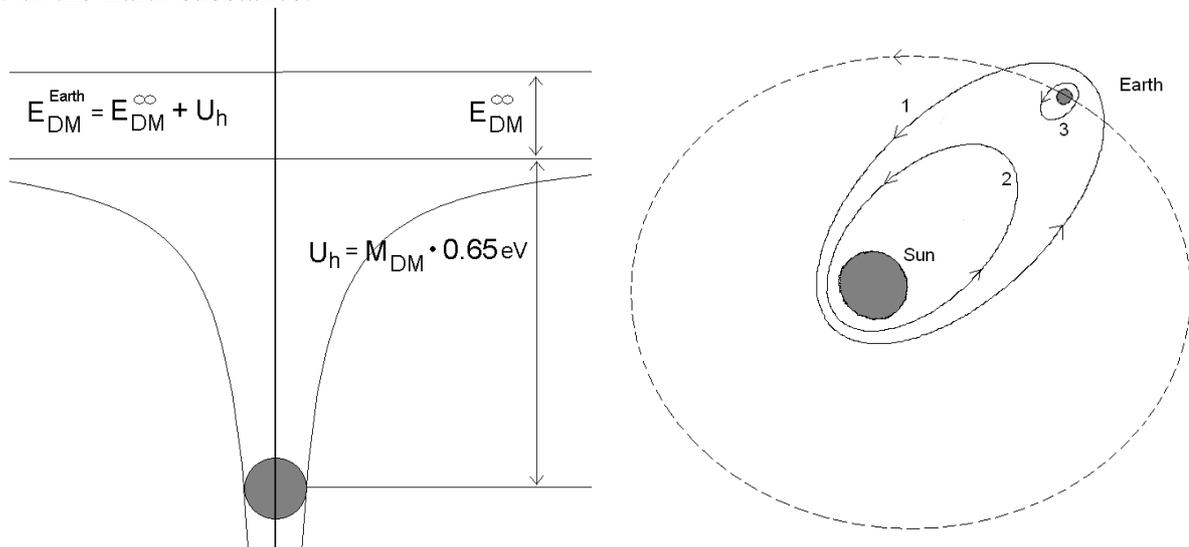

Fig. 5 a) Gravitational potential of the Earth. b) 1-trajectories of the particles captured by the system Sun-Earth. 2-trajectories of particles captured in a gravitational field of the Sun after energy loss at interaction with the Sun substance. 3-trajectories of particles captured in a gravitational field of the Earth after energy loss at interaction with substance of the Earth

Trajectories of the particles captured by the system Sun-Earth are shown in Fig. 5. These orbits (designated by figure 1) will be precessed around the Sun with frequency close to that of the Earth rotation around the Sun. As a result of interaction with substance of either the Sun or the Earth they will be captured in a gravitational hole of the Sun (transition to orbits 2) or in a gravitational hole of the Earth (transition to orbit 3). The density of dark matter particles in a gravitational hole of the Sun can be estimated according to the following scheme. Capture probability of dark matter particles in a gravitational hole of the Sun is equal $P_{capt}^{Sun} = n_{DM}^{Gal} \sigma_{Sun} v_{DM}$, where $\sigma_{Sun}$ - interaction cross - section of a dark matter with the sun. The dark matter particles density in a gravitational hole of the Sun can be estimated by very approximate formula $n_{DM}^{Sun} = P_{capt}^{Sun} \cdot T / V_{eff}^{Sun}$, where accumulation time $T \approx 10$ billion years, $V_{eff}^{Sun}$ - effective volume of a gravitational hole of the Sun. Unfortunately, detailed calculations have not been made. Very raw estimates show that the density of low-energy dark matter in a gravitational hole of the Sun would be by four order more than galactic density of a dark matter. The estimation of density of a dark matter in a gravitational hole of the Earth can be made by the same way.

The path length of dark matter particle at velocity of about 11-12 km/s is approximately 100 km (when $\alpha = 10^{26}$). Thus the capture cross section of a dark matter in a gravitational hole is equal to cross section of the Earth. An estimation of capture probability will be made for particles of dark matter with mass of 20 GeV $P_{capt}^{Earth} = 2 \cdot 10^{25} s^{-1}$. Here the factor of increasing dark matter particles density in a gravitational hole of the Sun is already taken into account. Consequently, the rate of increasing weight of the Earth is: $\Delta \dot{M} = 6.9 \cdot 10^2$ g/s. As a result, the increase of the Earth mass will be $2 \cdot 10^{20}$ g or $\Delta M/M = 3.4 \cdot 10^{-8}$ for 10 billion years. This estimation does not appear to contradict common sense. It is not an easy task to make an estimation of dark matter density on the Earth surface, since density distribution has not been calculated yet. A rough estimate can be made for average density in the volume of a sphere with $R = 10 R_{Earth}$ $\rho_{DM}^{average} = \frac{\Delta M}{V} = 2 \cdot 10^{-10}$ g/cm$^3$. Thus a rough estimate for the product of coefficients of increasing of low-energy dark matter density with respect to the galactic density can achieve about $10^{14} - 10^{15}$.

Certainly, the consideration given above is rather general. For an accurate calculation to be made one needs a computer modelling. The only conclusion to be drawn now is that the assumption on the long range character of interaction of dark matter particles with ordinary matter is not contradictory within the framework of the treatment involved. It is to be noted that there have been some restrictions on density of long-range dark matter on the Earth $\rho_{DM}^{E-surface} \leq 6.6 \cdot 10^{-13}$ g/cm$^3$ in the experiments with UCN. It is an important experimental information in the problem concerned.

Without finding obvious contradictions, we can continue considering the problem of possible application of UCN trap as a detector of long-range dark matter.

## 7. The possibility of observation of periodic variations of UCN storage time

A criterion for the dark matter signal can be a periodic variation of storage time in the experiment with UCN trap similar to that in the DAMA experiment. However, variations of the dark matter flux on the Earth surface where the UCN trap is located may be due to the reasons different from those of the DAMA experiment. The most basic assumptions about the behaviour of the dark matter captured by the system Earth-Moon are illustrated in Fig. 6. Dark matter is supposed to be localised on circumterraneous orbits near the Earth surface and moves at a velocity of approximately 8 km/s.

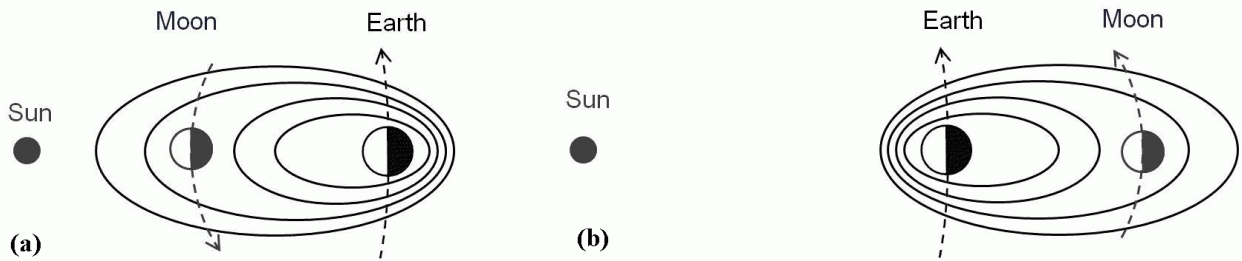

Fig. 6 Assumption of predicted behaviour of the dark matter captured by the system Earth-Moon

In Fig. 6 two phases associated with rotation of the Moon around the Earth are shown. The density of dark matter on the surface of the Earth at night is seen to be higher than that in the afternoon (Fig 6 a)). Therefore daily variations of the density of dark matter are expected to be observed with the detector, but the phase of daily variations will be changing all the time due to the rotation of the Moon around the Earth. For example, the phase changes by $180^o$ in position in Fig. 6 в) (full moon) in relation to the phase in Fig. 6 a) (new moon). Actually the period of variations will be 24 hours 50 minutes rather than 24 hours, since the Moon will appear over the detector exactly in accordance with this period.

The number of neutrons registered by means of a UCN detector after storage during the time $t_{hold}$ will be equal to: $N(t_{hold}) = N_0 e^{-t_{hold}\left(\tau_n^{-1} + \tau_{loss}^{-1} + \tau_{vac}^{-1} + \tau_{DM}^{-1}\right)} = A\left(1 + \dfrac{at_{hold}}{\tau_{DM}} \cos 2\pi\left(\dfrac{t-t_0}{T}\right)\right)$,

where $a$ - amplitude of variations of the density and the flux of dark matter, $T$ - cycle time of the Moon around a fixed point on the Earth surface - 24 hours 50 mines, $t_0$ - the phase connected with the position of the Moon on the orbit.

Certainly, the picture of the UCN count rate variations presented above can be more complicated if we take into account the influence of fluxes of dark matter due to the movement of the Earth along its orbit and its simultaneous rotation. However, the given treatment is based on the simplicity of the interpretation of the general idea, implying that the density of the dark matter captured by the system Earth-Moon is the highest one.

The consideration made at the end of the previous paragraph on trajectory attenuation in moving along orbits around the Earth and inside the Earth shows that a high gradient of the dark matter density should be expected at the sea level. Fluctuations of this dark matter level are likely to occur due to the gravitational influence on dark matter of the Moon and the Sun. Therefore the effects similar to sea tides might arise for dark matter, but they will have the opposite sign with respect to ocean high and low tides. If this analogy holds true, there should be observed two maxima within a day. Undoubtedly, this variation pattern will be different from that stated earlier.
Experiment performance on search of UCN storage time variations represents considerable interest.

## Conclusion

Within the bounds of the assumption that dark matter particles possess long-range interaction with $\lambda$ at orders of $10^{-5}$ cm, it is possible to draw the following conclusions.

1. Of considerable interest is the assumption on additional (except gravitational) character of interaction of dark matter particles and particles of ordinary matter at distances of an order of $10^{-5}$ cm, since neither experimental data nor theoretical models are available for the time being. Thus the first experimental restrictions obtained are of great significance.

2. The probability of dark matter influence on probability of the UCN storage does not exceed $2 \cdot 10^{-5} s^{-1}$ as it follows from comparison of the results on neutron lifetime. $/\left(\tau_n^{-1}\right)_{UCN} - \left(\tau_n^{-1}\right)_{beam} = \tau_{DM}^{-1} = (1.10 \pm 0.45) \cdot 10^{-5} s^{-1} < 2 \cdot 10^{-5} s^{-1}$ /

3. The probability of dark matter influence on heating UCN does not exceed $3 \cdot 10^{-7}$ s$^{-1}$ as it follows from the direct experiment on low-energy heating UCN at storage [14].

4. If one assumes that in the DAMA experiment the signal of long range dark matter has been observed, it is possible to set restriction on the upper limit for density of long range dark matter on the Earth: $\rho_{DM}^{E\text{-surface}} \leq 6.6 \cdot 10^{-13}$ g/cm$^{-3}$, using the experimental results with UCN.

5. The assumption concerned with long-range interaction ($\lambda \approx 10^{-5}$ cm) of dark matter particles and ordinary matter leads to a substantial enhancement of cross section of interaction with substance at the low energy. Consequently, there arises a possibility of capture and accumulation of dark matter in a gravitational field of the Earth. Rough estimation points to the possibility of accumulating low-energy dark matter on the Earth with density exceeding $10^{-13}$ g/cm$^3$. Computer modelling of the process of capture and accumulation of low-energy dark matter is of great necessity.

In conclusion one should say that preliminary estimations on the sensitivity of the experiment "UCN trap –DM detector" as compared with the DAMA experiment are not optimistic because of the low UCN density. However, introduction of long-range interactions ($\lambda \approx 10^{-5}$ cm) gives rise to the possibility of capturing and accumulating low-energy dark matter in a gravitational field of the Earth. In this case the method UCN is of great practical value since recoil energy at the level of few meV can be registered only in the experiment with UCN trap. The development of the proposed method seems to be reasonable as it opens a new field of studying low-energy dark matter possessing long-range forces.

## Acknowledgements

The authors are grateful for usefull discussions and critical remarks to Z. Berezhiani,
A.D. Dolgov, E. M. Drobyshevski, D.S. Gorbunov, V. Kauts, M.Yu. Khlopov, M.S. Lasakov, V.A. Rubakov and D.A. Varshalovich. The idea of this experimental method was arisen in time of visit at  Gran Sasso National Laboratory to Z. Berezhiani in February 2007. The first presentation of this work was done  9-th March  2010 on the XLIV PNPI Winter School. In the next article we are going to report the result of the first experiment search for daily variations of UCN storage time.

This work was supported by the Russian Foundation for Basic Research (project nos. 08-02-01052a,  10-02-00217a, 10-02-00224a)  and by the Federal Agency of Education of the Russian Federation (contract nos. P2427, P2540)  also by  the Federal Agency of Science and Innovations of the Russian Federation (conract  no. 02.740.11.0532).